\documentstyle[11pt]{article}
\begin{document}
\baselineskip 18pt
\begin{titlepage}
\centerline{\large\bf
 Determine the parameter $B_K$ of the $K^0-\overline{K^0}$ system
           }
\centerline{\large\bf
 by means of the precise Cabibbo-Kobayashi-Maskawa matrix elements
           }
\vspace{1cm}

\centerline{ Yong Liu$^{1,2}$, Jing-Ling Chen$^{2}$ and Wu-Sheng Dai$^3$ }
\vspace{0.5cm}
{\small
\centerline{\bf 1. Lab of Numerical Study for Heliospheric Physics}
\centerline{\bf  Chinese Academy of Sciences,
                 P. O. Box 8701, Beijing 100080, P.R.China}
\vspace{0.2cm}
\centerline{\bf 2. Theoretical Physics Division}
\centerline{\bf Nankai Institute of Mathematics}
\centerline{\bf Nankai University, Tianjin 300071, P.R.China}
\vspace{0.2cm}
\centerline{\bf 3. Department of Physics}
\centerline{\bf Nankai University, Tianjin 300071, P.R.China}
 }
\vspace{2cm}

\centerline{\bf Abstract}
\vspace{0.3cm}

Taking use of the relation between weak $CP$ phase
and the other three mixing angles in Cabibbo-Kobayashi-Maskawa (CKM)
matrix postulated by us before,
the uncertainty coming from the weak interaction can be
reduced, furthermore, by means of the relation between the calculation
result about mixing
parameter $\epsilon$ from the box diagrams and the related experimental
data in $K^0-\overline{K^0}$ system, the parameter $B_K$ can be
extracted.
We take $V_{ud},\; V_{ub}$ and $V_{tb}$ as inputs,
when we let them vary in the ranges
$
0.9745 \leq V_{ud} \leq 0.9760, \;\;
0.0018 \leq V_{ub} \leq 0.0045$ and $
0.9991 \leq V_{tb} \leq 0.9993
$,
we find the permitted window for $B_K$ is
$
0.444 \leq B_K \leq 1.242
$.
With the
more precise measurment on the CKM matrix elements in the future,
we can determine $B_K$ more precisely.   \\ \\
PACS number(s): 11.30.Er, 12.10.Ck, 12.39.Fe

\vspace{3cm}
\noindent
Email address: yongliu@ns.lhp.ac.cn
\end{titlepage}

\centerline{\large\bf
 Determine the parameter $B_K$ of the $K^0-\overline{K^0}$ system
           }
\centerline{\large\bf
 by means of the precise Cabibbo-Kobayashi-Maskawa matrix elements
           }
\vspace{1cm}

The heavy flavor physics has aroused a great interests in recent years
[1-4]. Through the study of heavy flavor physics,
we can extract much useful information. On one hand, the weak decays
of heavy flavours can provide a direct way to determine the weak mixing
angles and to test the unitarity of the standard Cabibbo-Kobayashi-Maskawa
(CKM) matrix [5-6].
On the other hand,
study on the semileptonic and
nonleptonic decays of heavy mesons can tell us some
information about non-perturbative QCD and its long range character.
But unfortunately, these two tasks are entangled.
When we want to determine
the weak mixing angles and the CP violation phase precisely, we must
know how to realiable evaluate hadronic matrix elements.
On the contrary, if we can
reduce some uncertainty coming from the weak interaction, we will
be able to determine
the relevant parameters of hadronic matrix element more precisely.

As a example, we consider the $K_0-\overline{K_0}$ system in this work.
Within some approximations, the CP violation parameter $\epsilon$
can be calculated [7-12]
\begin{equation}
|\epsilon|\simeq  \frac{G_F^2 m_K f_K^2 B_K M_W^2}{
  \sqrt{2} (12 \pi ^2) \Delta m_K} [\eta_1 S(x_c) I_{cc}+
  \eta_2 S(x_t) I_{tt}+2 \eta_3 S(x_c,x_t) I_{ct}]
\end{equation}
where $G_F$ is the Fermi constant,
$\eta_1=1.38, \eta_2=0.57, \eta_3=0.47$ are QCD corrections [8],
$I_{ij}\equiv Im(V_{id}^* V_{is} V_{jd}^* V_{js})$, $x_i=m_i^2/M_W^2$
and
\begin{equation}
S(x) \equiv \frac{x}{4}[1+\frac{3-9 x}{(x-1)^2}+\frac{6 x^2 ln(x)}{(x-1)^2}]
\end{equation}
\begin{equation}
S(x,y)\equiv x y \{[\frac{1}{4}+\frac{3}{2 (1-y)}-\frac{3}{4 (1-y)^2}]
\frac{ln(y)}{y-x}+(y\leftrightarrow x)-
\frac{3}{4 (1-x) (1-y)} \}.
\end{equation}

The experimental values of the relevant parameters are [9,13]
$$
|\epsilon|=(2.28\pm0.02)\times10^{-3} \;\;\;\; f_K=160 MeV \;\;\;\;
\Delta m_K=3.49\times10^{-15}GeV
$$
$$
m_K=0.4977 GeV \;\;\;\; m_t=175\pm6 GeV/c^2 \;\;\;\; M_W=80.34\pm0.10
GeV/c^2.
$$

The quantity $B_K$ should be in principle renormalization scale
independent, its value obtained by various groups with different methods
is listed in Table 1 [14-27].

\begin{table}[t]
\caption{ $B_K$ value obtain by various groups with different methods }
\begin{center}
\begin{tabular}{lll}
\hline
$B_K$ & Method & Ref. \\
\hline
3/4 & leading 1/$N_c$ &   [16] \\
0.37 & lowest-order chiral perturbation theory & [17]\\
0.70 $\pm$ 0.10 & next-to-leading 1/$N_c$ estimate & [18]\\
0.4 $\pm$ 0.2 & next-to-leading 1/$N_c$ estimate, $o(p^2)$ & [19]\\
0.42 $\pm$ 0.06 &$o(p^4)$ chiral perturbation theory & [20]\\
0.60-0.80 & NJL model with spin-1 interactions &[21]\\
0.39 $\pm$ 0.10 &QCD-hadronic duality &[22,23]\\
0.5$\pm$0.1 $\pm$0.2 &QCD sum rules (3-point functions) &[24]\\
0.55$\pm$0.25 &QCD sum rules (3-point functions) &[25]\\
0.58$\pm$0.22 & Laplace sum rule &[26]\\
0.90$\pm$0.03 $\pm$0.14 &lattice &[27]\\
\hline
\end{tabular}
\end{center}
\end{table}

From Table 1, we find that the value of $B_K$ is
very undefined. In fact, because
we are in ignorance of the low-energy QCD, we can not determine it
precisely. However, if we can understand the weak interaction very well,
through the study on the heavy flavor physics, conversely, we can
determine the relevant parameters of the strong interaction such as
$B_K$ more precisely.

The central purpose of this work is to determine $B_K$ by use of the
experimental results on $K_0-\overline{K_0}$ system and the CKM matrix
with the uncertainty coming from the weak CP phase $\delta$ being
reduced.

In Ref.[28], we find that the weak $CP$ phase is related to the other
three mixing angles, the relation can be described by
\begin{equation}
\sin\delta_{13}=\frac{ (1+s_{12}+s_{23}+s_{13})
                       \sqrt{1-s_{12}^2-s_{23}^2-s_{13}^2+
                       2 s_{12} s_{23} s_{13}} }{(1+
                       s_{12}) (1+s_{23}) (1+s_{13})}
\end{equation}
where $s_{ij}$ and $\delta_{13}$ are the parameters in the standard
parametrization [11-12]
\begin{equation}
V_{KM}= \left (
\begin{array}{ccc}
   c_{12} c_{13} & s_{12} c_{13}& s_{13} e^{-i \delta_{13}} \\
   -s_{12} c_{23}-c_{12} s_{23} s_{13} e^{i \delta_{13}} &
   c_{12} c_{23}-s_{12} s_{23} s_{13} e^{i \delta_{13}}    &
   s_{23} c_{13}\\
   s_{12} s_{23}-c_{12} c_{23} s_{13} e^{i \delta_{13}}  &
   -c_{12} s_{23}-s_{12} c_{23} s_{13} e^{i \delta_{13}} &
   c_{23} c_{13}
\end{array}
\right )
\end{equation}
with $c_{ij}=\cos\theta_{ij}$ and $s_{ij}=\sin\theta_{ij}$ for the
"generation" labels $i,j=1,2,3$.
Here, the real
angles $\theta_{12}, \theta_{23}$ and $\theta_{13}$ can all be made to
lie in the first quadrant. The phase $\delta_{13}$ lies in the range
$0<\delta_{13}<2 \pi$. In following, we will make the three angles
$\theta_{ij}$ lie in the first quadrant.

According to Eq.(4), the weak CP phase is fully determined by the three
mixing angles. Hence, the uncertainty coming from the weak interaction
is reduced. In following, we begin to determine $B_K$ with this result
is used. The program is

1. From three CKM matrix elements which are measured precisely, we
calculate the three mixing angles.

2. Based on Eq.(4), the weak CP phase $\delta_{13}$ can be solved. Hereon,
we can determine all the CKM matrix elements, and then,
we can calculate
$I_{ij} (i,j=c,t)$.

3. According to Eq.(1) and the relevant experimental results on
$K_0-\overline{K_0}$ system, we determine $B_K f_K^2$ or $B_K$.

The other parameters used in this work are
$$
G_F=1.166392\times10^{-5}GeV^{-2} \;\;\;\; m_c=1.5 GeV
$$
We take $V_{ud},\; V_{ub}$ and $V_{tb}$ as inputs,
When we let them vary in the ranges [13]
\begin{equation}
0.9745 \leq V_{ud} \leq 0.9760 \;\;\;\;\;\;
0.0018 \leq V_{ub} \leq 0.0045 \;\;\;\;\;\;
0.9991 \leq V_{tb} \leq 0.9993
\end{equation}
We find a range for $B_K$ as following
\begin{equation}
0.444 \leq B_K \leq 1.242
\end{equation}

However, if we let $V_{ud},\; V_{ub}$ and $V_{tb}$ vary in more
narrow ranges, such as
\begin{equation}
0.9749 \leq V_{ud} \leq 0.9756 \;\;\;\;\;\;
0.0025 \leq V_{ub} \leq 0.0038 \;\;\;\;\;\;
0.9991 \leq V_{tb} \leq 0.9993
\end{equation}
then, we find a range for $B_K$ as
\begin{equation}
0.520 \leq B_K \leq 0.902.
\end{equation}

In conclusion, by use of the postulation on the relation
between weak $CP$ phase and the other three mixing angles in CKM
matrix, the uncertainty coming from the weak interaction has been
reduced, furthermore, based on the relation between the calculation
result about mixing
parameter $\epsilon$ from the box diagrams and the related experimental
data in $K^0-\overline{K^0}$ system, the parameter $B_K$
is extracted.

We take $V_{ud},\; V_{ub}$ and $V_{tb}$ as inputs from the Data Book,
when all the three input parameters are on the $90\%$ CL, we find
$0.444 \leq B_K \leq 1.242$. So, it is consistent with most of the
results listed in Table 1, except those obtained by lowest-order chiral
perturbation theory and $o(p^4)$ chiral perturbation theory.

When the more narrow ranges for the inputs being scanned, we
will get a more narrow permitted window for $B_K$.
Hence, with the
more precise measurment on the CKM matrix elements in the future,
based on our postulation Eq.(4),
we will be able to determine $B_K$ more precisely.

\end{document}